\documentclass{webofc}
\usepackage[varg]{txfonts}   
\newcommand{\sNN}{s_\mathrm{NN}}
\newcommand{\dd}{\partial}
\begin{document}
\title{Study of Lambda polarization at RHIC BES and LHC energies}
%
%

\author{\firstname{Iurii} \lastname{Karpenko}\inst{1,2}\fnsep\thanks{\email{yu.karpenko@gmail.com}} \and
        \firstname{Francesco} \lastname{Becattini}\inst{1,3}\fnsep\thanks{\email{becattini@fi.infn.it}}
}

\institute{INFN - Sezione di Firenze, Via G. Sansone 1, I-50019 Sesto Fiorentino (FI), Italy
\and
SUBATECH, University of Nantes -- IN2P3/CNRS -- IMT Atlantique, Nantes, France
\and
Universit\'a di Firenze, Via G. Sansone 1, I-50019 Sesto Fiorentino (FI), Italy}

\abstract{%
In hydrodynamic approach to relativistic heavy ion collisions, hadrons with nonzero spin, produced out of the hydrodynamic medium, can acquire polarization via spin-vorticity thermodynamic coupling mechanism. The hydrodynamical quantity steering the polarization is the thermal vorticity, that is minus the antisymmetric part of the gradient of four-temperature field.

Based on this mechanism there have been several calculations in hydrodynamic and non-hydrodynamic models for non-central heavy ion collisions in the RHIC Beam Energy Scan energy range, showing that the amount of polarization of produced $\Lambda$ hyperons ranges from few percents to few permille, and decreases with collision energy. We report on an extension of our existing calculation of global $\Lambda$ polarization in UrQMD+vHLLE model to full RHIC and LHC energies, and discuss the component of polarization along the beam direction, which is the dominant one at high energies.
}
\maketitle
\section{Introduction}
\label{intro}
Recently, STAR collaboration has reported on the first observation of significantly nonzero global polarization of $\Lambda$($\bar\Lambda$) hyperons produced in non-central Au-Au collisions in the RHIC Beam Energy Scan (BES) Program \cite{STAR:2017ckg}.
The magnitude of measured polarization is in general agreement with calculations from different hydrodynamic models \cite{Karpenko:2016jyx,Xie:2017upb} and AMPT \cite{Li:2017slc, Sun:2017xhx}. In the hydrodynamic models the $\Lambda$ hyperons produced at particlization (fluid to particle transition) hypersurface acquire polarization via spin-vorticity coupling mechanism. In case of AMPT the spin-vorticity coupling mechanism is introduced with the help of coarse-graining procedure. The magnitude of polarization depends on local {\it thermal vorticity} of the fluid, $\varpi_{\mu\nu}=(\dd_\nu \beta_\mu - \dd_\mu \beta_\nu)$ at the points of $\Lambda$ production, $\beta_\mu=u_\mu/T$ being the inverse four-temperature field. In this proceeding we report on an extension of our existing calculations of global $\Lambda$ polarization in RHIC BES \cite{Karpenko:2016jyx} and discuss the polarization component along beam direction.

\section{Model and results}
We study production and polarization of $\Lambda$ in a hybrid transport model, which consists of initial state UrQMD hadron/string cascade, followed by a 3 dimensional viscous hydrodynamic expansion. At a hypersurface where the local rest frame energy density reaches a critical value $\epsilon_{\rm sw}=0.5$~GeV/fm$^3$, the particlization (i.e.\ transition from fluid to particle description) is performed with the help of Cooper-Frye formula. Along with the rest of hadrons, $\Lambda$ as well as other heavy reasonances which decay to it, are produced and acquire their polarization at this hypersurface. Details of the cascade+viscous hydrodynamic model used are thoroughly described in \cite{Karpenko:2015xea}.

We have reported global polarization of $\Lambda$ hyperons calculated with this model in the energy range $\sqrt{\sNN}=7.7,\dots,200$~GeV in \cite{Karpenko:2016jyx}. It was concluded that, in line with the experimental results, in 20-50\% central Au-Au collisions the global $\Lambda$ polarization decreases with increasing collision energy from about 2\% at $\sqrt{\sNN}=7.7$~GeV to 0.2\% at $\sqrt{\sNN}=200$~GeV. Only about one quarter of $\Lambda$ are the ones produced directly at particlization, the rest are products of resonance decays, including cascade processes $X\rightarrow\Sigma^0\rightarrow\Lambda$. We have estimated \cite{Becattini:2016gvu} that such resonance feed-down results in 15\% reduction of the polarization of all $\Lambda$ at all collision energies in consideration.

\begin{figure*}[h]
\centering
\includegraphics[width=0.49\textwidth,clip]{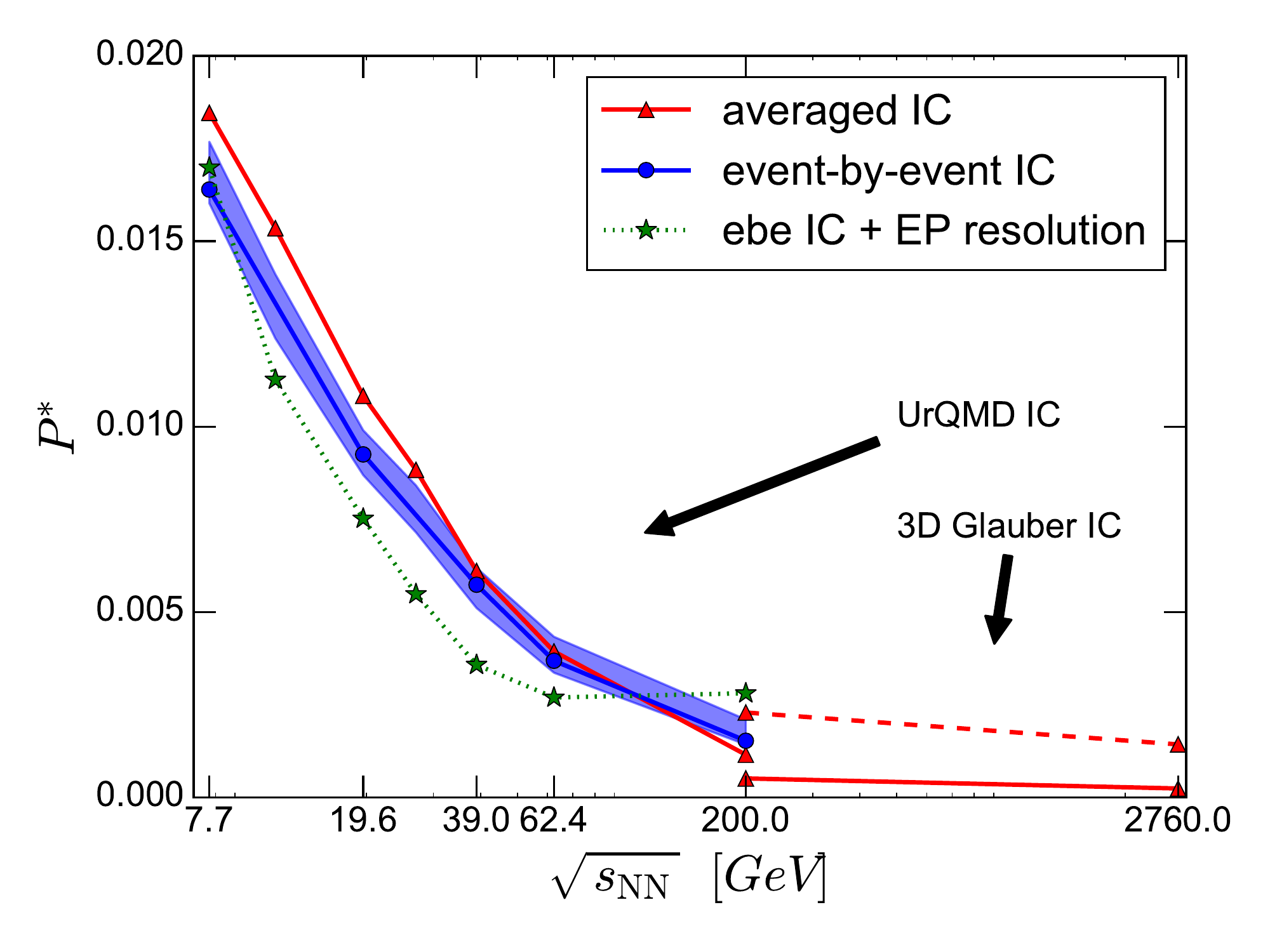}
\includegraphics[width=0.49\textwidth,clip]{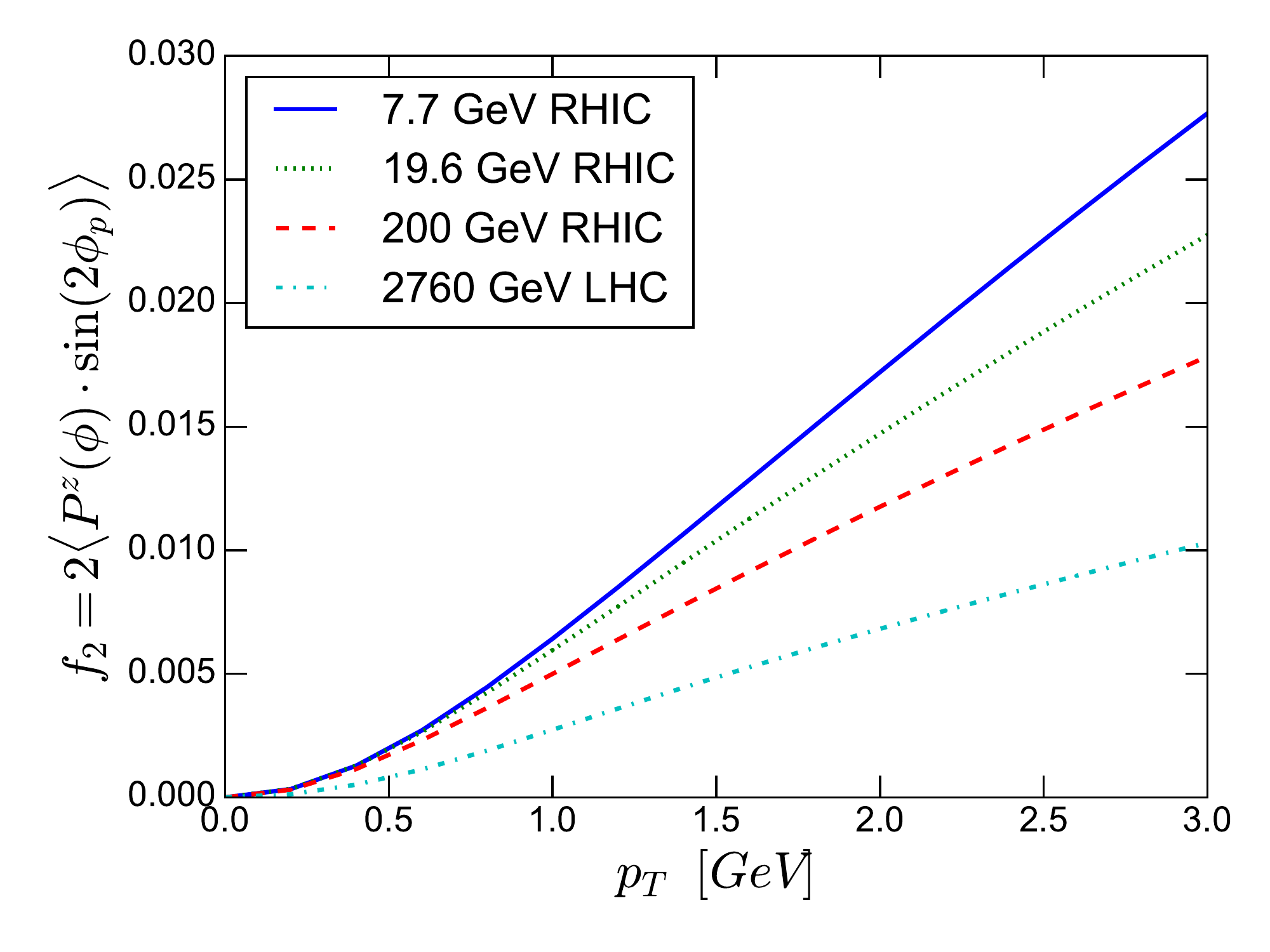}
\caption{Left: collision energy dependence of global $\Lambda$ polarization along global angular momentum of the fireball, calculated with averaged initial state from UrQMD ($\sqrt{\sNN}$=7.7,\dots,200 GeV) and Monte Carlo Glauber ($\sqrt{\sNN}$=200,2760 GeV). Right: second order Fourier harmonic coefficien of polarization component along beam direction, calculated as a function of $p_T$ for different collision energies; 200 and 2760 GeV points correspond to Monte Carlo Glauber initial state.}
\label{fig1}       
\end{figure*}

We have extended this result with a 3+1 dimensional viscous hydrodynamic calculation for $\sqrt{\sNN}=200$ GeV RHIC Au-Au and 2760 GeV Pb-Pb collisions using a Monte Carlo Glauber initial state with parametrized rapidity dependence \cite{Bozek:2012fw} tuned to describe $dN/dy$ and directed flow of charged hadrons at these energies. The combined results are shown on Fig.~\ref{fig1}, left. For the Glauber IS, we show two lines (connecting $\sqrt{\sNN}=$200 and 2760 GeV points), the solid one corresponding to longitudinally boost invariant initial flow, and the dashed one corresponding to a small amount of initial shear longitudinal flow as described in \cite{Becattini:2015ska}. The latter results in higher initial thermal vorticity which in turn leads to a larger magnitude of $\Lambda$ polarization. Also, for the RHIC BES range we show the curves for both averaged initial state and event-by-event hydro with fluctuating initial state. From the plot one can conclude that 1) the global polarization signal in the model is relatively insensitive to whether the initial state is fluctuating or not, and 2) the global $\Lambda$ polarization further decreases with energy from RHIC to LHC, and although there are no published results yet from the LHC, we estimate that at the LHC the global polarization possibly goes below the experimental limit for being detected.

One expects that the global ({\it i.e.}\ integrated over all $\Lambda$ in an event) polarization vector is directed parallel to the one of global angular momentum, which is confirmed by our calculations. However doing a $p_T$ differential analysis we have found that, at high energies, for mid-rapidity $\Lambda$ the dominant component of  polarization is the one along the beam direction \cite{Becattini:2017gcx}. In a simplified scenario with no initial state fluctuations, this component has a quadrupole structure in the transverse momentum plane, and can be decomposed into Fourier series with only sine terms of even multiples of the azimuthal angle $\phi$ of the transverse momentum vector:
\begin{equation}\label{szfour}
 P^z({\bf p}_T,Y=0) = \sum_{k=1}^\infty f_{2k}(p_T) \sin 2 k \varphi.
\end{equation}
In this expansion, the dominant component is $f_2$, which we show on the right panel of Fig.~\ref{fig1}. Contrary to global polarization (left panel of the same figure) which decreases by almost 1 order of magnitude between $\sqrt{\sNN}$=7.7 and 200 GeV, $f_2$ decreases by only 35\%. Even at 2.76~TeV LHC energy, it reaches 1\% for $p_T=3$~GeV $\Lambda$, making it feasible to detect it in the experiment.

Another useful feature of the $f_{2k}$ components of the Fourier expansion is that, different from the global polarization vector, they are independent on the orientation (sign) of the event plane, which simplifies the experimental measurements.

The structure of the polarization component along the beam direction can also be accessed via correlations of the corresponding polarization components of $\Lambda$ pairs. One can show with a simple derivation that, keeping the second Fourier harmonic only:

\begin{equation}
P^z=f_2\sin2(\phi-\Psi)\quad\Rightarrow\quad \langle P^z(\phi)P^z(\phi+\Delta\phi)\rangle_\phi=
{f_2^2 \over 2}\cos2\Delta\phi,
\end{equation}

where $\langle A\rangle_\phi={1\over 2\pi}\int A(\phi)d\phi$. Note that such covariance function is independent of the event plane angle $\Psi$.

\begin{figure*}
\centering
\includegraphics[width=0.49\textwidth,clip]{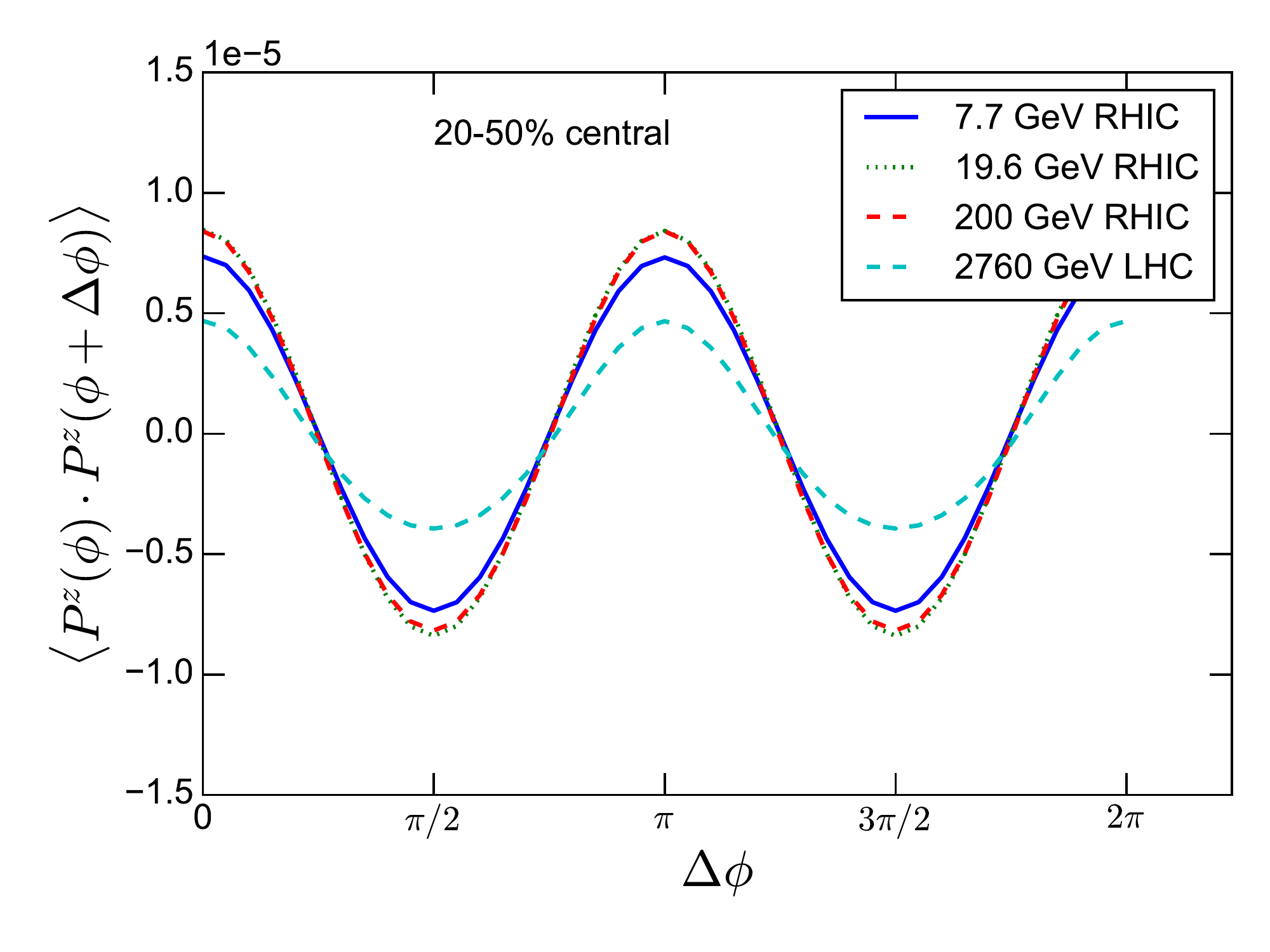}
\includegraphics[width=0.49\textwidth,clip]{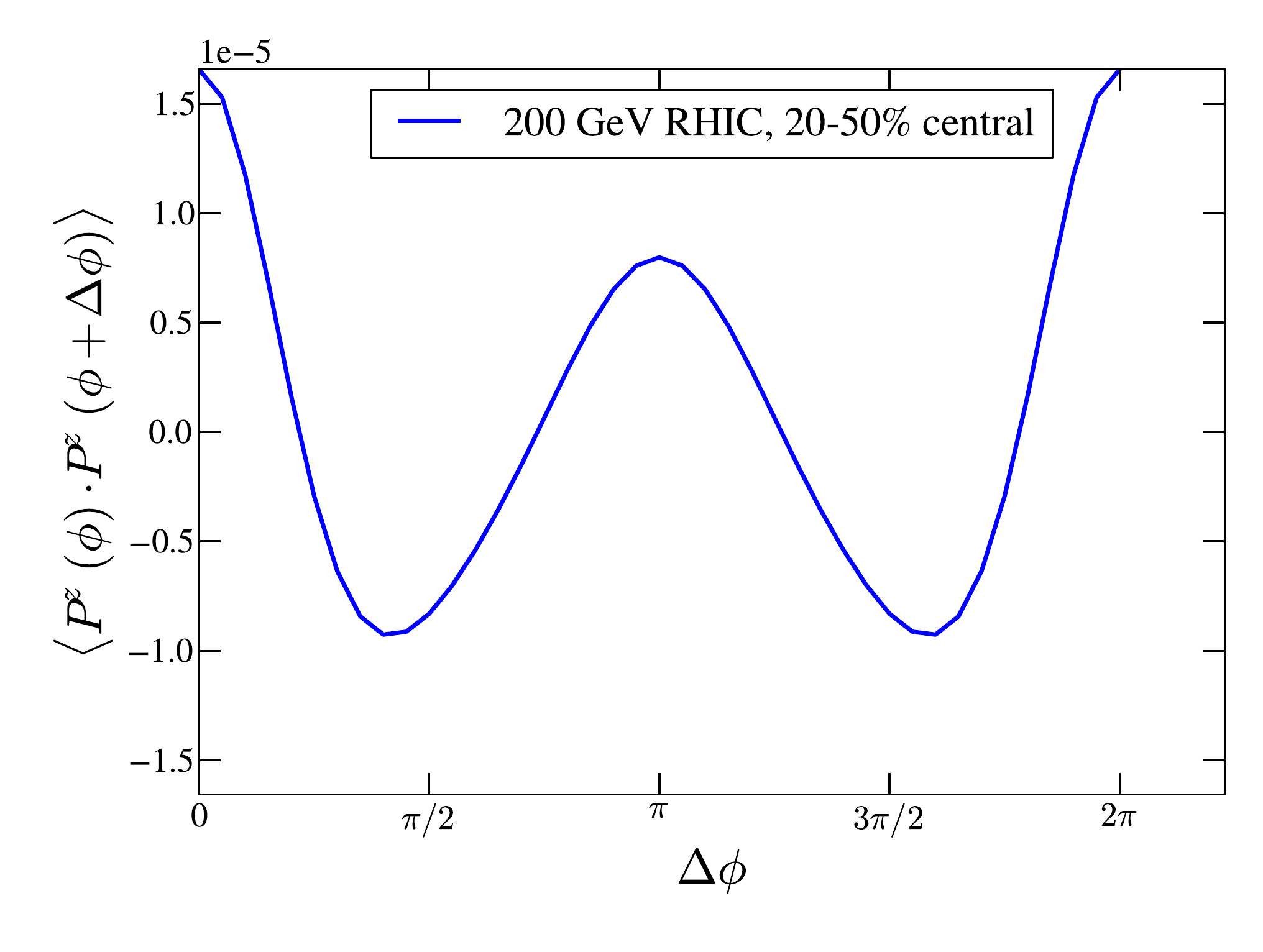}
\caption{Covariance of polarization components in beam direction, for pairs of midrapidity $\Lambda$ separated by azimuthal angle $\Delta\phi$ in transverse momentum space. Right panel: results with averaged initial state from UrQMD ($\sqrt{\sNN}$=7.7,\dots,200 GeV Au-Au) and Monte Carlo Glauber ($\sqrt{\sNN}$=200 Au-Au, 2760 GeV Pb-Pb). Left panel: event-by-event hydrodynamic calculation with UrQMD initial state for Au-Au collisions at $\sqrt{\sNN}=200$~GeV.}
\label{fig2}       
\end{figure*}

The resulting covariance functions for $P^z$ are plotted on Fig.~\ref{fig2}. The left panel corresponds to averaged initial state and a single hydrodynamic simulation for given collision energy, where due to the geometrical symmetry and the above argument we obtain the expected $\cos(2\Delta\phi)$ shape, with collision energy dependent amplitude. The right panel shows the covariance function calculated in event-by-event hydrodynamic case for $\sqrt\sNN=200$~GeV RHIC scenario. For the event-by-event case, we observe a relative enhancement of the covariance function at $\Delta\phi=0$, caused by vorticity patterns formed by hot spots in the initial state. The overall shape and magnitude of the covariance function are qualitatively consistent with the results in \cite{Pang:2016igs}.

Note that the quadrupole pattern of the longitudinal component of polarization is present in case of average (non-fluctuating) initial state. Even more, this effect is still present and has a similar magnitude when the initial state has a complete longitudinal boost invariance. The reason is that the longitudinal component of polarization is linked to the anisotropies in the transverse expansion, therefore the only requisite for it is anisotropic transverse flow, which also drives the elliptic flow. Indeed we have shown that in a simple Blast-Wave model there is a linear relation between the $p_T$ dependent $f_2$ harmonic and elliptic flow:
\begin{equation}
   f_2(p_T) = 2 \frac{d T}{d \tau}\frac{1}{mT} v_2(p_T)
\end{equation}
where $dT/d\tau$ is assumed time derivative of temperature $T$ (which in Bjorken picture depends only on proper time $\tau$) at the freezeout hypersurface, and $m$ is mass of particle.

\section{Conclusions}

We have extended the results for global $\Lambda$ polarization in RHIC Beam Energy Scan \cite{Karpenko:2016jyx} by using Monte Carlo Glauber initial state with parametrized rapidity dependence, tuned to reproduce basic observables at $\sqrt\sNN=$200 GeV RHIC and 2760 GeV LHC energies. We find that, as expected, the global polarization of $\Lambda$ hyperons further decreases from the full RHIC to the LHC energy. However, at such high energies the dominant component of $\Lambda$ polarization vector in $p_T$ differential analysis is the one along the beam direction. This component of polarization has a quadrupole structure in the transverse momentum plane, and shows a mild decrease with collision energy. Its second order Fourier harmonic for $p_T=3$~GeV $\Lambda$ reaches 1\% at 2760 GeV LHC energy. Such quadrupole structure also results in correlations of polarization components of $\Lambda$ pairs. We stress that the longitudinal component of polarization is a generic effect which emerges even in a simple longitudinal boost invariant non-fluctuating hydrodynamic picture of heavy ion collisions.

\begin{acknowledgement} {\bf Acknowledgements}: This work was partly supported by the University of Florence
grant {\it Fisica dei plasmi relativistici: teoria e applicazioni moderne}.
\end{acknowledgement}

%
%
%

\end{document}